\documentclass[aps,prd,twocolumn,superscriptaddress,preprintnumbers]{revtex4-2}

\usepackage{amssymb}
\usepackage{amsmath}
\usepackage{graphicx}
\usepackage{enumerate}
\usepackage{mathtools}
\usepackage{color}
\usepackage{bbold}
\usepackage{subfigure}
\usepackage{slashed}
\usepackage[dvipsnames,usenames]{xcolor}
\usepackage{bm}
\usepackage{ulem}
\usepackage{centernot}

\begin{document}

% Use the \preprint command to place your local institutional report
% number in the upper righthand corner of the title page in preprint mode.
% Multiple \preprint commands are allowed.
% Use the 'preprintnumbers' class option to override journal defaults
% to display numbers if necessary

\preprint{LA-UR-23-34030}

%Title of paper
\title{Collisional flavor pendula and neutrino quantum thermodynamics}

% repeat the \author .. \affiliation  etc. as needed
% \email, \thanks, \homepage, \altaffiliation all apply to the current
% author. Explanatory text should go in the []'s, actual e-mail
% address or url should go in the {}'s for \email and \homepage.
% Please use the appropriate macro foreach each type of information

% \affiliation command applies to all authors since the last
% \affiliation command. The \affiliation command should follow the
% other information
% \affiliation can be followed by \email, \homepage, \thanks as well.

\author{Lucas Johns}
\email[]{ljohns@lanl.gov}
\affiliation{Theoretical Division, Los Alamos National Laboratory, Los Alamos, NM 87545, USA}
\affiliation{Department of Physics, University of California, Berkeley, CA 94720, USA}

\author{Santiago Rodriguez}
\affiliation{Department of Physics, University of California, Berkeley, CA 94720, USA}

\begin{abstract}
Neutrinos in core-collapse supernovae and neutron-star mergers are susceptible to flavor instabilities of three kinds: slow, fast, and collisional. Prior work has established mappings of the first two onto abstract mechanical systems in flavor space, respectively named the slow and fast flavor pendula. Here we introduce and analyze the flavor pendulum associated with the third class. We explain our results in terms of the recently developed theory of neutrino quantum thermodynamics. Perhaps our most surprising finding is that there exists a limit in which decoherent interactions drive perfectly coherent flavor conversion.
\end{abstract}

\maketitle

\section{Introduction}

In core-collapse supernovae and neutron-star mergers, neutrino propagation is influenced by coherent self-interactions \cite{fuller1987, notzold1988, pantaleone1992, pantaleone1995, qian1995a, sigl1995, pastor2002b, friedland2003, balantekin2005}. From this process arise collective flavor-oscillation phenomena whose astrophysical effects are yet to be worked out.

One way to approach the problem is to study flavor instabilities, which are indicative of discrepancies between neutrino transport with and without oscillations \cite{banerjee2011}. Instabilities fall into three classes: slow \cite{kostelecky1993b, duan2006c}, fast \cite{sawyer2005, sawyer2016, chakraborty2016c}, and collisional \cite{johns2023}. Our knowledge of the conditions and consequences of all three is rapidly progressing \cite{tamborra2017, dasgupta2017, wu2017, wu2017b, abbar2019b, azari2019, nagakura2019, martin2020, johns2020, johns2020b, bhattacharyya2020, bhattacharyya2021, shalgar2020b, xiong2020, morinaga2020, glas2020, abbar2020b, george2020, johns2021, nagakura2021d, nagakura2021b, nagakura2021c, richers2021c, li2021, padillagay2022, richers2022b, harada2022, dasgupta2022, capozzi2019, johns2022, just2022, fernandez2022, nagakura2023, shalgar2023d, shalgar2023e, myers2022, grohs2023, grohs2023b, ehring2023b, ehring2023, bhattacharyya2022, johns2022b, shalgar2021, kato2021, sasaki2022, hansen2022b, padillagay2022b, kato2022b, kato2023, lin2023, padillagay2022b, fiorillo2023c, xiong2023b, xiong2023, liu2023, xiong2023c, nagakura2023c, shalgar2023b, zaizen2023, zaizen2023b, liu2023b, fischer2023, kato2023b, johns2023c}. 

Analytic results are generally hard to obtain in this research area because of the nonlinearity introduced by self-interactions. Those that can be found are valuable. They inform our overall understanding of oscillation physics and motivate useful approximations of various kinds. The cardinal results of this nature are a set of mappings from idealized models of neutrino gases to simple mechanical systems. Insofar as the latter are tractable and intuitive, they elucidate the more opaque physics of dense neutrino populations.

The pioneering contributions in this spirit established the formal equivalence between an isotropic, homogeneous neutrino gas and a gyroscopic pendulum \cite{hannestad2006, duan2007b, duan2007c, raffelt2011}. The latter system is now known as the \textit{slow flavor pendulum} because its fall from an inverted position corresponds to the simplest instance of slow flavor instability. Subsequent work has expanded the analysis to include anisotropy \cite{raffelt2007b, raffelt2013} and inhomogeneity \cite{mangano2014}. The slow pendulum also pertains to neutrino flavor evolution in the early universe, though its dynamics is qualitatively distinct from what is generally expected in supernovae and mergers \cite{johns2016, johns2018}. 

Fast instability maps onto the motion of the \textit{fast flavor pendulum} under the conditions of axial symmetry in momentum and either homogeneity in space or stationarity in time \cite{johns2020}. The fast pendulum has been used to formulate approximate stability criteria \cite{johns2020, johns2021}, to predict the extent of flavor conversion \cite{bhattacharyya2020, padillagay2022, bhattacharyya2022}, to interpret the numerical successes and limitations of angular-moment methods \cite{johns2020b, myers2022, grohs2023}, to understand the modification of fast flavor conversion by collisions \cite{shalgar2021, kato2021, sasaki2022, hansen2022b, johns2022, padillagay2022b, kato2022b, kato2023}, and to construct soliton solutions for the neutrino flavor field \cite{fiorillo2023c}. Both the slow and the fast pendulum illuminate the results of many-body neutrino calculations, which commonly employ models whose mean-field approximations are flavor pendula or integrability-broken versions thereof \cite{pehlivan2011, xiong2022c, martin2023, martin2023b, fiorillo2023, shalgar2023c, johns2023b, patwardhan2023, balantekin2023b}.

These mechanical equivalences are mathematically rich because they involve the emergence of simple, integrable dynamics from continuous distributions of collectively interacting particles \cite{hannestad2006, johns2020, padillagay2022, fiorillo2023, xiong2023b}. From a physical standpoint, they are like the analytic building blocks of collective neutrino oscillations. 

Given the existence of pendulum solutions for slow and fast instabilities, an obvious question is whether such a solution exists for collisional instabilities. In this paper we show that an essential mechanism of collisional instability is indeed the motion of the \textit{collisional flavor pendulum}. 

The slow and fast pendula move in the presence of gravitational fields set up by neutrino mass splitting \cite{hannestad2006} and neutrino spectral asymmetry \cite{johns2020, padillagay2022}, respectively. In the simplest analysis, the collisional pendulum experiences no gravity. Instability is the result of energy exchange with the environment. The outcome of this exchange---the particular evolution that results from the instability---depends on whether the spin of the pendulum about its own axis is initially right-handed ($\sigma > 0$), left-handed ($\sigma < 0)$, or vanishing ($\sigma = 0$). A $\sigma < 0$ attractor comes to dominate the evolution sooner or later. Collisional flavor conversion is a matter of how the system gets to the attractor, and what happens once it does.

Under typical conditions, the collision rate of neutrinos exceeds that of antineutrinos ($\Gamma > \bar{\Gamma}$) due to the neutron-richness of the medium. Assuming this and the additional property that the number densities of the heavy-lepton species are all comparable, the three dynamical regimes above correspond to situations where electron neutrinos outnumber electron antineutrinos ($n_{\nu_e} > n_{\bar{\nu}_e}$), where the opposite ordering holds ($n_{\nu_e} < n_{\bar{\nu}_e}$), and where the densities are the same ($n_{\nu_e} = n_{\bar{\nu}_e}$). The importance of the number-density ordering for collisional instability is known from past studies \cite{johns2023, johns2022b, lin2023, padillagay2022b, xiong2023, liu2023, xiong2023c, nagakura2023c, shalgar2023b, liu2023b, fischer2023, kato2023b}. For instance, resonance-like collisional instabilities, which exhibit growth rates proportional to $G_F^{3/2}$ as opposed to the standard $G_F^2$, occur where $n_{\nu_e} \approx n_{\bar{\nu}_e}$ \cite{xiong2023c}. Collisional flavor swaps have been observed in the same region of parameter space \cite{kato2023b}.

For collisional instabilities, it would be reasonable to think that the analogue mechanical system surely cannot be a conservative one. Collisional decoherence, which is the cause of collisional instability, is irreversible. Yet we find here that in fact the collisional pendulum can be arbitrarily close to energy- and length-conserving even as it completely inverts its position. In other words, there is a limit in which the flavor evolution is coherent despite being driven by decoherent interactions. This is a discovery regarding the dynamical possibilities of open quantum systems in general, and its implications potentially extend well beyond neutrino physics.

Our analysis is based on coarse-graining. The same concept is the foundation of neutrino quantum thermodynamics (\textit{i.e.}, the theory of mixing equilibrium) and miscidynamics (\textit{i.e.}, the theory of neutrino transport predicated on local mixing equilibrium) \cite{johns2023c}. The findings we report can be neatly accounted for by these theories, and in that sense provide further support for them.

After defining our model system (Sec.~\ref{sec:model}) and reviewing its linearized behavior (Sec.~\ref{sec:linear}), we present the pendulum analysis summarized above (Sec.~\ref{sec:pendulum}). We close with some brief remarks about the possibility of coherent collisional evolution and about the relationship between this study and neutrino quantum thermodynamics (Sec.~\ref{sec:discussion}).

\section{The model \label{sec:model}}

In this paper we restrict our attention to the very simple system consisting of an isotropic, homogeneous, single-energy, two-flavor neutrino gas undergoing coherent self-interactions and collisional decoherence. We neglect vacuum oscillations, coherent scattering on the matter background, and all collisional effects beyond coherence damping.

The flavor composition of the neutrinos is contained in the density matrix $\rho = \left( P_0 + \boldsymbol{P} \cdot \boldsymbol{\sigma} \right) / 2$, which we expand in Pauli matrices to obtain the polarization vector $\boldsymbol{P}$. Antineutrino symbols carry bars (\textit{e.g.}, $\boldsymbol{\bar{P}}$). The equations of motion are
\begin{gather}
\dot{\boldsymbol{P}} = \mu \left( \boldsymbol{P} - \boldsymbol{\bar{P}} \right) \times \boldsymbol{P} - \Gamma \boldsymbol{P}_T, \notag \\
\dot{\boldsymbol{\bar{P}}} = \mu \left( \boldsymbol{P} - \boldsymbol{\bar{P}} \right) \times \boldsymbol{\bar{P}} - \bar{\Gamma} \boldsymbol{\bar{P}}_T, \label{eq:basicEOMs}
\end{gather}
where $\mu P_z = \sqrt{2} G_F \left( n_{\nu_e} - n_{\nu_x} \right)$. $T$ denotes the part of the vector transverse to the flavor axis $\boldsymbol{z}$. The traces $P_0$ and $\bar{P}_0$ are constant.

Rather than working with the polarization vectors directly, we introduce the sum and difference vectors $\boldsymbol{S} \equiv \boldsymbol{P} + \boldsymbol{\bar{P}}$ and $\boldsymbol{D} \equiv \boldsymbol{P} - \boldsymbol{\bar{P}}$, for which the equations of motion are
\begin{gather}
\dot{\boldsymbol{S}} = \mu \boldsymbol{D} \times \boldsymbol{S} - \Gamma_+ \boldsymbol{S}_T - \Gamma_- \boldsymbol{D}_T, \notag \\
\dot{\boldsymbol{D}} = - \Gamma_+ \boldsymbol{D}_T - \Gamma_- \boldsymbol{S}_T. \label{eq:SDEOMs}
\end{gather}
We are defining $\Gamma_+ \equiv \left( \Gamma + \bar{\Gamma} \right) / 2$ and $\Gamma_- \equiv \left( \Gamma - \bar{\Gamma} \right) / 2$. Note that this differs from the notation of Ref.~\cite{johns2023}, where the $+$ and $-$ subscripts were used to signify collisional asymmetries between the flavors rather than between neutrinos and antineutrinos.

Following Ref.~\cite{johns2023}, we observe that neglecting the oscillation term $\mu \boldsymbol{D} \times \boldsymbol{S}$ leads to the equations
\begin{equation}
\ddot{\boldsymbol{S}}_T + \Gamma_+ \dot{\boldsymbol{S}}_T - \Gamma_-^2 \boldsymbol{S}_T \cong 0 \label{eq:oldSinst}
\end{equation}
and
\begin{equation}
\dot{\boldsymbol{D}}_T \cong \left( \pm \frac{S}{D} \Gamma_- - \Gamma_+ \right) \boldsymbol{D}_T, \label{eq:oldDinst}
\end{equation}
where $S \equiv | \boldsymbol{S} |$ and $D \equiv | \boldsymbol{D} |$. We assume that $n_{\nu_x} = n_{\bar{\nu}_x}$, which implies that $D_z = n_{\nu_e} - n_{\bar{\nu}_e}$. The unstable solutions of Eqs.~\eqref{eq:oldSinst} and \eqref{eq:oldDinst} then correspond to the collisional instability in the regimes $n_{\nu_e} > n_{\bar{\nu}_e}$ and $n_{\nu_e} < n_{\bar{\nu}_e}$, respectively. The apparent singular behavior of Eq.~\eqref{eq:oldDinst} as $D \rightarrow 0$ hints at the existence and enhanced growth rate of the resonance-like collisional instability \cite{xiong2023c}. Because $D \cong D_z$ during the linear phase of evolution, the resonance-like regime has $n_{\nu_e} \approx n_{\bar{\nu}_e}$.

The argument given in Ref.~\cite{johns2023} for dropping the oscillation term is that $\boldsymbol{S}$ and $\boldsymbol{D}$ synchronize, \textit{i.e.}, maintain (anti)parallel orientations. In this work we refine the argument by asserting synchronization of the vectors after temporally coarse-graining the dynamics. This allows us to go beyond Ref.~\cite{johns2023} and gain analytic insight into the \textit{nonlinear} evolution when $n_{\nu_e} \not\approx n_{\bar{\nu}_e}$. Furthermore, coarse-graining is consonant with Ref.~\cite{johns2023c}, where it was used to make the passage from neutrino quantum kinetics to neutrino quantum thermodynamics. Through our analysis here we also get deeper insight into the survival of collisional instabilities in the miscidynamic equation---which is a coarse-grained, or grid-level, description of neutrino transport---in contrast with the absorption of other oscillation phenomena into the condition of local mixing equilibrium \cite{johns2023c}.

As we will see, resonance-like collisional instability stands apart from the collisional instabilities symbolized by Eqs.~\eqref{eq:oldSinst} and \eqref{eq:oldDinst} in that it is poorly approximated by synchronization even at the coarse-grained level. That is why it fails to show up in Eq.~\eqref{eq:oldDinst} except through the suggestion of a singularity. Fine-grained details are responsible for regularizing the behavior at $D_z \approx 0$---which is to say, they are responsible for the instability itself. The distinctive $G_F^{3/2}$ growth rate reflects a failure of scale separation between terms at order $G_F$ (\textit{i.e.}, coherent self-interactions) and at order $G_F^2$ (\textit{i.e.}, collisions).

As this paper was in final editing, another study \cite{fiorillo2023d} appeared that uses scale separation and time-averaging in the analysis of collisional instabilities.

\section{Linear stability analysis \label{sec:linear}}

In this section we present the main points of the linear stability analysis of our chosen system.

We take Eqs.~(13) of Ref.~\cite{johns2023} as our starting point. Setting the vacuum oscillation frequency to zero, we have
\begin{align}
i \partial_t \rho_{ex} &= \left( - \sqrt{2} G_F ( n_{\bar{\nu}_e} - n_{\bar{\nu}_x} )  - i \Gamma  \right) \rho_{ex} \notag \\
& \hspace{0.25 in} + \sqrt{2}G_F ( n_{\nu_e} - n_{\nu_x} ) \bar{\rho}_{ex}, \notag \\
i \partial_t \bar{\rho}_{ex} &= \left( + \sqrt{2} G_F ( n_{\nu_e} - n_{\nu_x} )  - i \bar{\Gamma}  \right) \bar{\rho}_{ex} \notag \\
& \hspace{0.25 in} - \sqrt{2}G_F ( n_{\bar{\nu}_e} - n_{\bar{\nu}_x} ) \rho_{ex}. \label{rhoex}
\end{align}
To find the system's collective modes, we adopt the ansatz
\begin{equation}
\rho_{ex} = Q e^{-i \Omega t}, ~~ \bar{\rho}_{ex} = \bar{Q} e^{-i \Omega t}.
\end{equation}
The pair $( Q, \bar{Q} )$ is the eigenvector associated with eigenvalue $\Omega$. After the ansatz is applied and Eqs.~\eqref{rhoex} are combined, $Q$ and $\bar{Q}$ drop out. We are left with the dispersion relation
\begin{align}
\Omega^2  + \left( -\mu D_z + 2 i \Gamma_+ \right) \Omega - i \mu& \left( D_z \Gamma_+ + S_z \Gamma_- \right) \notag \\
&- \Gamma_+^2 + \Gamma_-^2 = 0, \label{disprel}
\end{align}
whose solutions are the complex frequencies of the collective modes.

We now let $\Omega \equiv \nu + i\gamma$, with $\nu$ and $\gamma$ real, and write down separate relations for the real and imaginary parts of Eq.~\eqref{disprel}:
\begin{gather}
\nu^2 - \gamma^2 - \mu D_z \nu - 2 \Gamma_+ \gamma - \Gamma_+^2 + \Gamma_-^2 = 0, \notag \\
2 \nu \gamma - \mu D_z \gamma + 2 \Gamma_+ \nu - \mu \left( D_z \Gamma_+ + S_z \Gamma_- \right) = 0. \label{reimeqtns}
\end{gather}
The lower line implies the following equation relating $\nu$ to $\gamma$:
\begin{equation}
\nu = \frac{\mu}{2} \left( D_z + \frac{\Gamma_-}{\gamma + \Gamma_+} S_z \right). \label{gammatonu}
\end{equation}
Plugging this into the upper line of Eq.~\eqref{reimeqtns}, assuming that
\begin{equation}
|\gamma| ~ \sim ~ \Gamma, \bar{\Gamma} ~ \ll ~ \mu |S_z|, \mu |D_z|, \label{sinsthierarchy}
\end{equation}
and retaining only the highest-order terms in $\mu$, we obtain
\begin{equation}
\gamma = - \Gamma_+ + \frac{S_z}{D_z} \Gamma_-. \label{gammasinst}
\end{equation}
Appealing to Eq.~\eqref{gammatonu}, we find the real part of this mode to be $\nu = \mu D_z$. Our findings are summarized by
\begin{equation}
\Omega_R = \mu D_z + i \left( - \Gamma_+ + \frac{S_z}{D_z} \Gamma_- \right), \label{eq:OmegaR}
\end{equation}
where the subscript $R$ anticipates the identification of $\Omega_R$ with the right-handed pendulum. Given our definition of $\Omega$, the mode is unstable when $\textrm{Im}~\Omega_R > 0$. Since we are assuming $\Gamma > \bar{\Gamma}$, instability occurs when $D_z$ and $S_z$ have the same sign. In the next section we will see that the handedness of the pendulum is given by $\textrm{sgn} (D_z S_z)$.

The unstable mode of the left-handed pendulum can be obtained by returning to the lower line of Eq.~\eqref{reimeqtns} and solving for $\gamma$ in terms of $\nu$:
\begin{equation}
\gamma = - \frac{\Gamma_+ + \frac{S_z}{D_z} \Gamma_- - 2 \frac{\nu}{\mu D_z}\Gamma_+}{1 - 2 \frac{\nu}{\mu D_z}}.
\end{equation}
For $| \nu | \ll \mu | D_z |$, $\gamma$ is approximately independent of $\nu$, with a value
\begin{equation}
\gamma = - \Gamma_+ - \frac{S_z}{D_z} \Gamma_-.
\end{equation}
Note that $D_z$ appears here with the opposite sign relative to Eq.~\eqref{gammasinst}. A distinct instability is indeed being revealed. Using the upper line of Eq.~\eqref{reimeqtns}, we arrive at
\begin{equation}
\nu = \frac{\Gamma_+^2 - \Gamma_-^2}{\mu D_z}.
\end{equation}
Collecting these expressions, we have
\begin{equation}
\Omega_L = \frac{\Gamma_+^2 - \Gamma_-^2}{\mu D_z} + i \left( - \Gamma_+ - \frac{S_z}{D_z} \Gamma_- \right). \label{eq:OmegaL}
\end{equation}
This mode is unstable when $\textrm{sgn}(D_z S_z) = -1$.

\begin{figure}
\centering
\includegraphics[width=.40\textwidth]{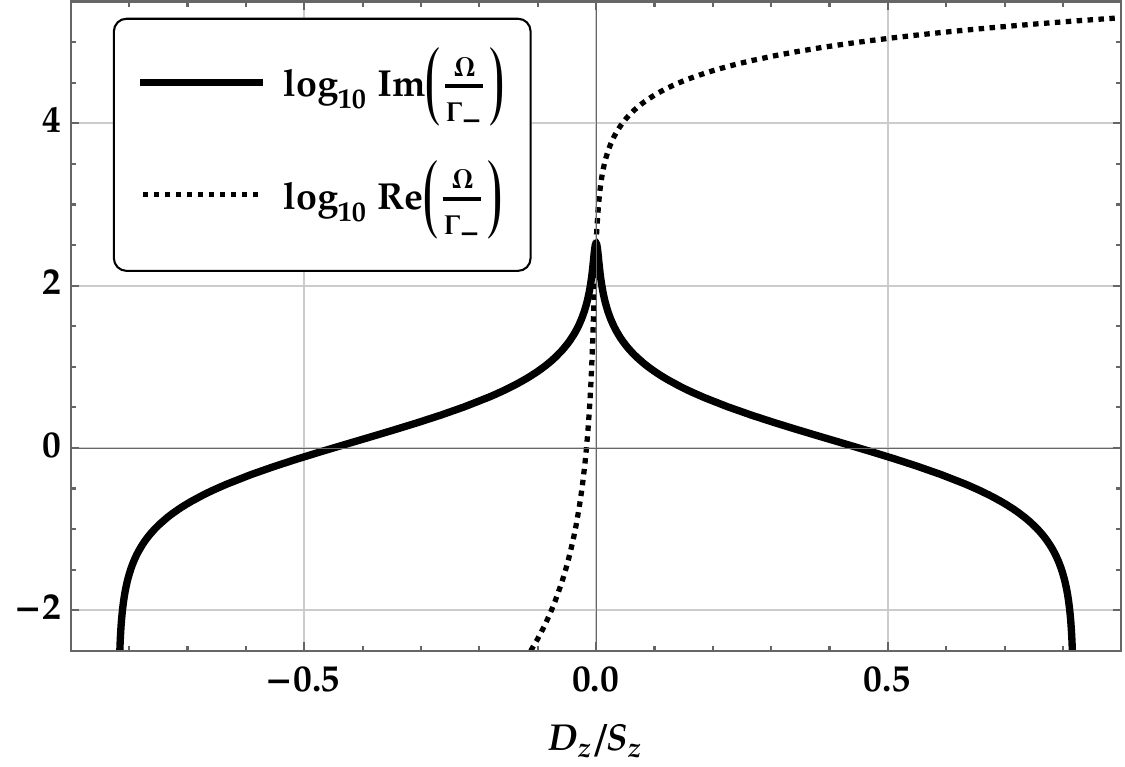}
\caption{Growth rate (solid) and oscillation frequency (dashed) of the collisional flavor instability, \textit{i.e.}, the unstable solution of Eq.~\eqref{disprel}. The chosen parameter ratios are $\Gamma / \bar{\Gamma} = 10$ and $\mu S_z / \bar{\Gamma} = 10^6$. In this example, the resonance-like ($D_z / S_z \cong 0$) instability achieves a maximum growth rate 330 times larger than $\Gamma_-$.}
\label{mode_vs_dz}
\end{figure}

Lastly we consider $D_z = 0$. Drawing on Eqs.~\eqref{reimeqtns}, the real part of the collective mode satisfies the quartic equation
\begin{align}
\gamma^4 &+ 4 \Gamma_+ \gamma^3 + \left( 6 \Gamma_+^2 - \Gamma_-^2 \right) \gamma^2 + 2 \Gamma_+ \left( 2 \Gamma_+^2 - \Gamma_-^2 \right) \gamma \notag \\
&+ \Gamma_+^2 \left( \Gamma_+^2 - \Gamma_-^2 \right) - \left( \frac{\mu S_z \Gamma_-}{2} \right)^2 = 0. \label{tinstpoly}
\end{align}
The final term on the lefthand side suggests a possible solution
\begin{equation}
| \gamma | \sim \mu | S_z \Gamma_- |
\end{equation}
in the limit
\begin{equation}
\mu |S_z| \gg \Gamma, \bar{\Gamma}.
\end{equation}
Compare to Eq.~\eqref{sinsthierarchy}, where $\mu | D_z |$ was large as well. Here, keeping only the first and last terms on the lefthand side of Eq.~\eqref{tinstpoly}, we get
\begin{equation}
\gamma = - \Gamma_+ + \sqrt{\frac{\mu | S_z \Gamma_- |}{2}},
\end{equation}
which is consistent with our assumption regarding the magnitude of $\gamma$. Using Eq.~\eqref{gammatonu} to calculate the real part, we find
\begin{equation}
\nu = \sqrt{\frac{\mu | S_z \Gamma_- |}{2}}.
\end{equation} 
Putting the pieces together, we have the mode associated with the zero-spin pendulum:
\begin{equation}
\Omega_0 = \sqrt{\frac{\mu | S_z \Gamma_- |}{2}} + i \left( - \Gamma_+ + \sqrt{\frac{\mu | S_z \Gamma_- |}{2}} \right). \label{eq:Omega0}
\end{equation}
Solving the full quartic equation for $\gamma$, then expanding the solution, leads to the same result.

To help visualize the results of this section, we plot in Fig.~\ref{mode_vs_dz} the numerical solution of the dispersion relation for a specific set of parameters. A similar figure is presented in Ref.~\cite{xiong2023c}.

\section{The pendulum \label{sec:pendulum}}

We now perform an analysis of the nonlinear dynamics exhibited by Eqs.~\eqref{eq:basicEOMs}.

To begin, we define the coarse-graining operator
\begin{equation}
\left\langle f (t) \right\rangle \equiv \frac{1}{\tau} \int_{t-\frac{\tau}{2}}^{t+\frac{\tau}{2}} d t' ~f( t' )
\end{equation} 
acting on function $f(t)$. We also introduce notation for the coarse-grained sum and difference vectors:
\begin{equation}
\boldsymbol{Q} (t) \equiv \left\langle \boldsymbol{S} (t) \right\rangle, ~~ \boldsymbol{L} (t) \equiv \left\langle \boldsymbol{D} (t) \right\rangle.
\end{equation}
The motivation for coarse-graining is to take advantage of the large scale separation between the self-interaction potential and the collision rate. We therefore choose a coarse-graining window $\tau$ somewhere in between the disparate scales:
\begin{equation}
\left( G_F n_\nu \right)^{-1} \ll \tau \ll \left( G_F^2 E_\nu^2 n_b \right)^{-1},
\end{equation}
where $n_\nu$ is the overall neutrino density, $E_\nu$ is a typical neutrino energy, and $n_b$ is the baryon density. We aim to find equations for $\boldsymbol{Q}$ and $\boldsymbol{L}$ such that the coarse-grained dynamics is influenced by the existence---but not the details---of the fine-grained variables. This is the same approach to oscillation physics that has been advanced in Refs.~\cite{johns2020, johns2020b, johns2023c}.

Upon coarse-graining Eqs.~\eqref{eq:SDEOMs}, we immediately encounter the question of what to do about the term $\mu \left\langle \boldsymbol{D} \times \boldsymbol{S} \right\rangle$. We build our analysis on the hypothesis that, for any vectors $
\boldsymbol{U}$ and $\boldsymbol{V}$ in flavor space,
\begin{gather}
\left\langle \boldsymbol{U} \times \boldsymbol{V} \right\rangle \cong \left\langle \boldsymbol{U} \right\rangle \times \left\langle \boldsymbol{V} \right\rangle, ~~
\left\langle \boldsymbol{U} \cdot \boldsymbol{V} \right\rangle \cong \left\langle \boldsymbol{U} \right\rangle \cdot \left\langle \boldsymbol{V} \right\rangle. \label{eq:ergodic}
\end{gather}
In other words, we ignore all fine-grained correlations. The relationship between this assumption and the notion of local mixing equilibrium \cite{johns2023c} is discussed in Sec.~\ref{sec:discussion}. Using Eqs.~\eqref{eq:ergodic}, we obtain the coarse-grained equations of motion:
\begin{gather}
\dot{\boldsymbol{Q}} = \mu \boldsymbol{L} \times \boldsymbol{Q} - \Gamma_+ \boldsymbol{Q}_T - \Gamma_- \boldsymbol{L}_T, \notag \\
\dot{\boldsymbol{L}} = - \Gamma_+ \boldsymbol{L}_T - \Gamma_- \boldsymbol{Q}_T. \label{eq:QLEOMs}
\end{gather}
Note that $L_z = D_z = \textrm{constant}$.

Now let us ask: Is there a limit in which $\boldsymbol{Q}$ behaves like a gyroscopic pendulum?

It is clear that pendulum motion, if it occurs, cannot persist indefinitely. As $t \rightarrow \infty$ the system must converge to the fixed point at $\boldsymbol{L} = L_z \boldsymbol{z}$, $\boldsymbol{Q} = 0$. But while $\boldsymbol{Q}$ depolarization is inevitable, it remains a possibility that its extent can be made arbitrarily small over a \textit{finite} duration. 

Heartened by this logic, we suppose that
\begin{equation}
\left\langle \boldsymbol{S}_T \right\rangle \cong 0, ~~ \left\langle D_T \right\rangle \cong 0, ~~ \left\langle S \right\rangle \cong \textrm{constant}  \label{eq:STDTzero}
\end{equation}
throughout some finite period of evolution beginning at $t = 0$. We do not assume that $\left\langle S_T \right\rangle$ is close to zero. Doing so would disallow instability of $\boldsymbol{S}$.
These conditions lead to the equation of motion
\begin{equation}
\dot{\boldsymbol{Q}} = \mu \boldsymbol{L} \times \boldsymbol{Q} \label{eq:LtimesQ}
\end{equation}
with $\boldsymbol{L} = L_z \boldsymbol{z} = \textrm{constant}$ and $Q = \left\langle S \right\rangle = \textrm{constant}$. As hoped for, there is no depolarization. Defining
\begin{equation}
\boldsymbol{q} \equiv \boldsymbol{Q} / Q, ~~ \zeta \equiv \boldsymbol{L} \cdot \boldsymbol{q},
\end{equation}
we obtain
\begin{equation}
\boldsymbol{L} = \frac{\boldsymbol{q} \times \dot{\boldsymbol{q}}}{\mu} + \zeta \boldsymbol{q}
\end{equation}
and, because $\zeta$ is implied to be constant by Eq.~\eqref{eq:LtimesQ} and the line beneath it,
\begin{equation}
\frac{\boldsymbol{q} \times \ddot{\boldsymbol{q}}}{\mu} + \zeta \dot{\boldsymbol{q}} = 0. \label{eq:sigmapend}
\end{equation}
This is the equation of motion of a gyroscopic pendulum in flavor space with unit length and spin $\zeta$ \cite{hannestad2006}. It is not quite the equation we are after, however. 

There is one more step for us to take. From Eqs.~\eqref{eq:SDEOMs} and \eqref{eq:STDTzero} we derive
\begin{align}
\frac{d}{dt} \left\langle \boldsymbol{D} \cdot \boldsymbol{S} \right\rangle &= - 2 \Gamma_+ \left\langle \boldsymbol{D}_T \cdot \boldsymbol{S}_T \right\rangle - \Gamma_- \left( \left\langle S_T^2 \right\rangle + \left\langle D_T^2 \right\rangle \right) \notag \\
&= - \Gamma_- \left\langle S_T^2 \right\rangle. \label{eq:zetaEOM}
\end{align}
Furthermore, Eqs.~\eqref{eq:ergodic} lead to
\begin{equation}
\zeta \cong \frac{\left\langle \boldsymbol{D} \cdot \boldsymbol{S} \right\rangle}{Q}.
\end{equation}
Based on these results, we make the replacement $\zeta \rightarrow \sigma(t)$ in Eq.~\eqref{eq:sigmapend}, where
\begin{equation}
\sigma (t) \equiv \frac{\left\langle \boldsymbol{D} \cdot \boldsymbol{S} \right\rangle (t)}{Q}. \label{eq:sigmadefn}
\end{equation}
In this definition, $\left\langle \boldsymbol{D} \cdot \boldsymbol{S} \right\rangle (t)$ is understood to be a solution of Eq.~\eqref{eq:zetaEOM}. We thus obtain
\begin{equation}
\frac{\boldsymbol{q} \times \ddot{\boldsymbol{q}}}{\mu} + \sigma (t) \dot{\boldsymbol{q}} = 0, \label{eq:pendulum}
\end{equation}
the equation of motion of the collisional flavor pendulum $\boldsymbol{q}$. Unlike the slow and fast pendula, this one experiences no gravity. Any interesting motion must arise from the secular evolution of the spin.

\begin{figure}
\centering
\includegraphics[width=.42\textwidth]{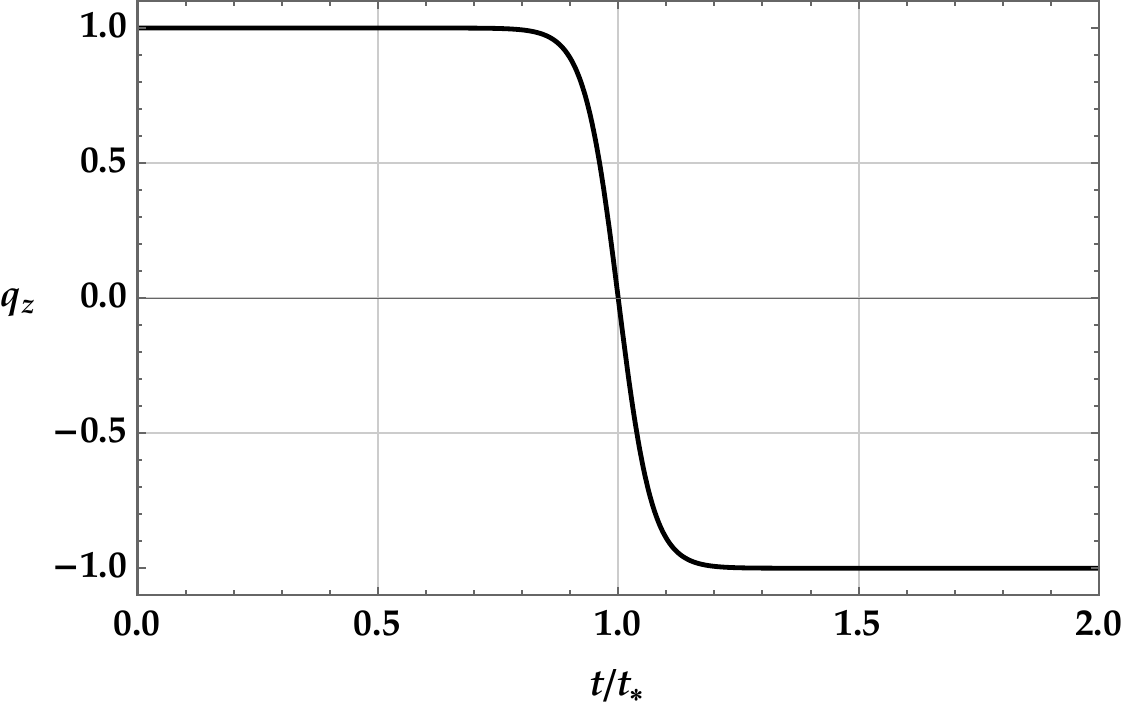}
\caption{Height $q_z$ of the collisional flavor pendulum according to the solution in Eq.~\eqref{eq:riccatisoln}. Time is in units of $t_*$ from Eq.~\eqref{eq:tstar}. The initial height is $q_z (0) = 1 - 10^{-12}$. The angular momentum is constant with value $\boldsymbol{L} = L_z \boldsymbol{z}$, $L_z > 0$.}
\label{riccati}
\end{figure}

Adiabatic invariance of angular momentum $\boldsymbol{L}$ implies that the height $q_z (t)$ of the pendulum is controlled entirely by $\sigma (t)$. Using Eqs.~\eqref{eq:ergodic} and \eqref{eq:zetaEOM}, we see that
\begin{equation}
\frac{d \left\langle S_z \right\rangle}{dt} = - \Gamma_- \frac{\left\langle S_T \right\rangle^2}{D_z}
\end{equation}
and therefore
\begin{equation}
\dot{q}_z = - \frac{Q \Gamma_-}{L_z}  \left( 1 - q_z^2 \right).
\end{equation}
The latter is a Riccati equation. Taking $q_z (0) = 1 - \delta$, the solution is
\begin{equation}
q_z (t) =  \frac{2 - \delta \left( 1 + \exp \frac{2 Q \Gamma_- t}{L_z} \right)}{2 - \delta \left( 1 - \exp \frac{2 Q \Gamma_- t}{L_z} \right)}. \label{eq:riccatisoln}
\end{equation}
The time at which $q_z$ crosses through zero is
\begin{equation}
t_* = \frac{L_z}{2 Q \Gamma_-} \log \frac{2-\delta}{\delta}. \label{eq:tstar}
\end{equation}
The pendulum height can alternatively be written in terms of a logistic function:
\begin{equation}
q_z(t) = 1 - \frac{2}{1 + \exp \left( \frac{2 Q \Gamma_-}{L_z} \left( t - t_* \right) \right)}.
\end{equation}
The solution is plotted in Fig.~\ref{riccati} using $\sigma (0) > 0$ and a value of $10^{-12}$ for $\delta$. As the pendulum drops from a height of $+1$ to $-1$, angular momentum is converted from spin to orbital and then back to spin. The orbital motion around $\boldsymbol{z}$ has frequency $\mu D_z$. Note the consistency with the right-handed collective mode whose frequency $\Omega_R$ is given in Eq.~\eqref{eq:OmegaR}. 

Unexpectedly, the dynamics we are observing in connection with collisional instability is similar to the spin-reversal mechanism of the neutrino flavor pendulum in the early universe \cite{johns2018}. Here, collisional decoherence is the source of fine-grained irreversibility and the driver of the dynamics. In the cosmological setting, Hubble expansion is the source and driver. In Sec.~\ref{sec:discussion} we interpret the coherent (polarization-preserving) inversion of the collisional flavor pendulum as an adiabatic process. Hubble expansion during the relevant cosmic epoch is also adiabatic. 

\begin{figure}
\centering
\includegraphics[width=.42\textwidth]{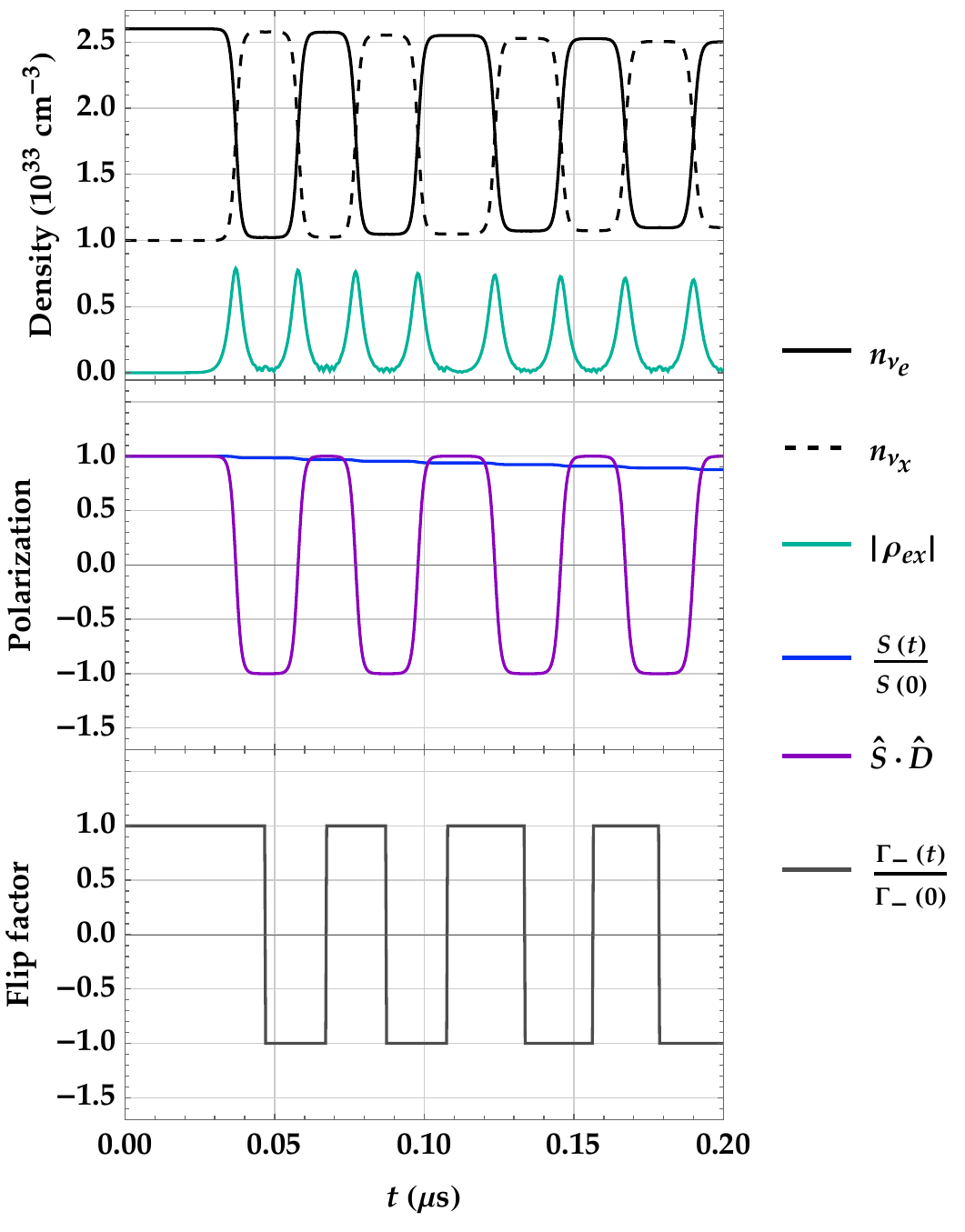}
\caption{Repeated occurrence of the nearly coherent collisional flavor instability. For this calculation, the sign of $\Gamma_-$ is artificially reversed each time $| \mathbf{\hat{S}} \cdot \mathbf{\hat{D}} |$ grows past the threshold of $1 - \epsilon_\textrm{flip}$, where $\epsilon_\textrm{flip} = 5 \times 10^{-5}$. A more diligent execution of this idea would result in less depolarization and more exact periodicity. Due to the choice of initial conditions, $n_{\bar{\nu}_e}$ differs from $n_{\nu_e}$ only by a fraction of a percent and is therefore not shown.}
\label{s_inst_only}
\end{figure}

After inverting from its initial position, the pendulum has left-handed spin and is stable according to Eq.~\eqref{eq:riccatisoln}. Nonetheless, nearly periodic motion is possible in principle because the inversion can occur with arbitrarily small changes in the length and energy of the pendulum. Figure~\ref{s_inst_only} demonstrates this point. In the numerical calculation, the sign of $\Gamma_-$ reverses whenever $| \mathbf{\hat{S}} \cdot \mathbf{\hat{D}} |$ comes within $\epsilon_\textrm{flip}$ of unity. This scenario has no relation to any astrophysical setting. The purpose is merely to highlight the surprising finding that in a certain limit (see Sec.~\ref{sec:discussion}) decoherent interactions drive perfectly coherent evolution.

Under more realistic conditions, we know from Sec.~\ref{sec:linear}, and in particular from $\Omega_L$ in Eq.~\eqref{eq:OmegaL}, that in fact the pendulum has not reached a stable position after inverting. Moreover, we know from $\Omega_0$ in Eq.~\eqref{eq:Omega0} that a pendulum with $\sigma (0) = 0$ is unstable as well. Equation~\eqref{eq:riccatisoln} does not adequately describe either of these instabilities.

\begin{figure*}
\centering
\begin{subfigure}{
\centering
\includegraphics[width=.45\textwidth]{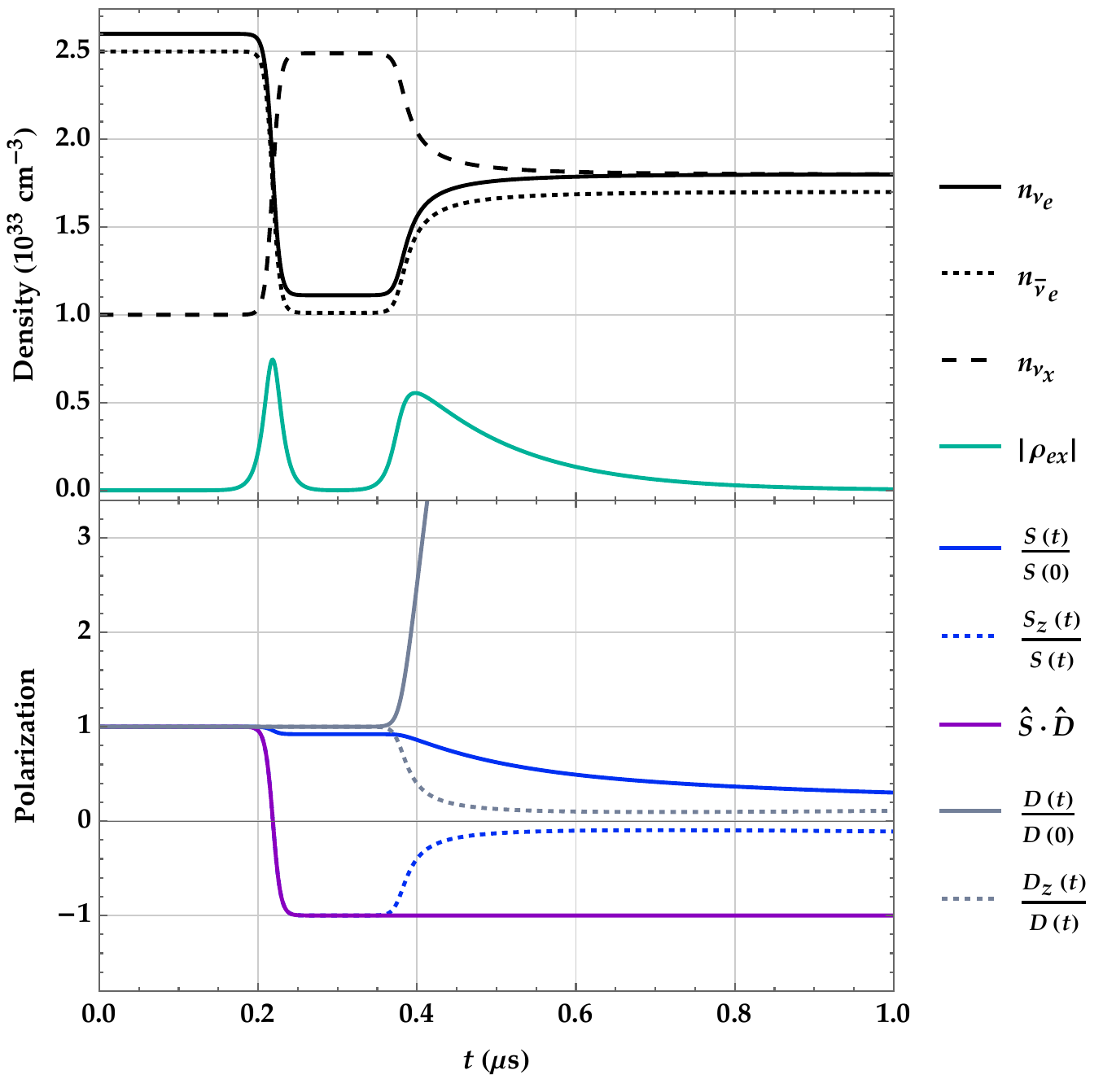}
}
\end{subfigure}\hspace{.25 in}
\begin{subfigure}{
\centering
\includegraphics[width=.44\textwidth]{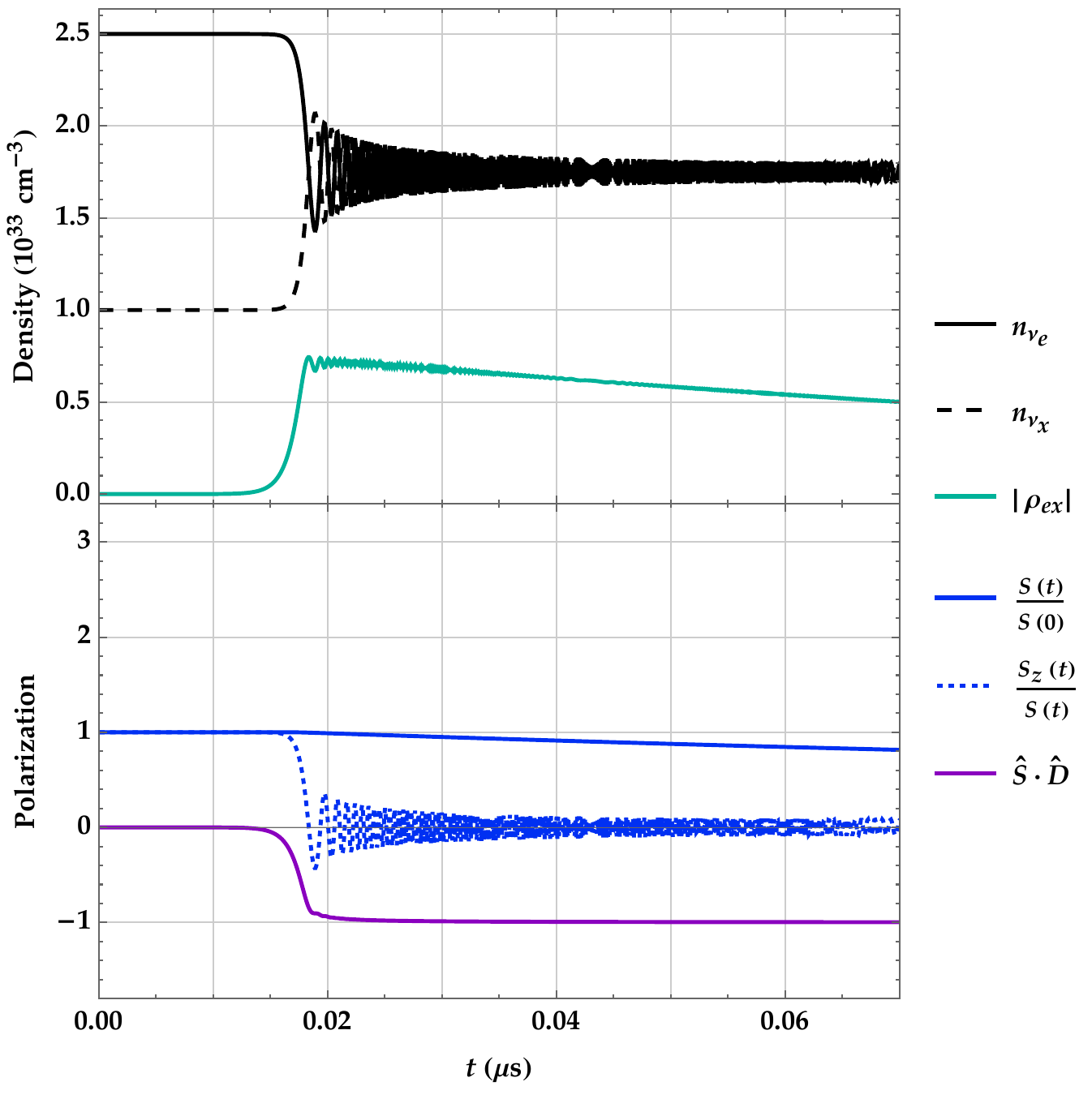}
}
\end{subfigure}
\caption{\textit{Left panel:} The initially right-handed ($\sigma > 0$) neutrino system. An inversion of the collisional flavor pendulum first brings the system to a metastable position. The left-handed ($\sigma < 0$) collisional instability then sets in and the flavor distributions asymptote to fully depolarized states. \textit{Right panel:} The initially zero-spin ($\sigma = 0$) neutrino system. Note the different scales of the horizontal axes. See text for discussion.}
\label{numerical}
\end{figure*}

To proceed, we return to our foundational Eqs.~\eqref{eq:ergodic}. We can actually guess that there are three qualitative regimes by noting that there are three ways for $\left\langle \boldsymbol{D} \times \boldsymbol{S} \right\rangle \cong \left\langle \boldsymbol{D} \right\rangle \times \left\langle \boldsymbol{S} \right\rangle$ to be true. The first is dephasing. There is significant fine-grained motion of $\boldsymbol{S}$ relative to $\boldsymbol{D}$, but it averages out under coarse-graining. This is the situation we have analyzed thus far. Per Eq.~\eqref{eq:STDTzero}, we have $\left\langle \boldsymbol{S}_T \right\rangle \cong 0$ due to phase cancellations within the time-averaging window.

The second way is to have $\boldsymbol{\hat{D}} \cong \pm \boldsymbol{\hat{S}}$ at all times, so that there is minimal fine-grained motion of $\boldsymbol{S}$ relative to $\boldsymbol{D}$. We have already seen how an initial configuration with $\boldsymbol{\hat{D}} \cong + \boldsymbol{\hat{S}}$ is unstable and inverts to the opposite orientation. But $\boldsymbol{\hat{D}} \cong - \boldsymbol{\hat{S}}$ brings us to the qualitatively distinct instability in Eq.~\eqref{eq:oldDinst}, or to
\begin{equation}
\dot{\boldsymbol{L}}_T \cong \left( \frac{Q}{L} \Gamma_- - \Gamma_+ \right) \boldsymbol{L}_T \label{eq:spinup}
\end{equation}
in terms of the coarse-grained variables. The environment injects mechanical energy into the pendulum. Because $\boldsymbol{Q}$ remains parallel to $\boldsymbol{L}$, the energy transfer takes the form of a spinning-up of $\sigma$. At the same time that the angular momentum $L$ is growing, the pendulum length $Q(t) / Q(0)$ is decaying. Eventually the $\Gamma_+$ term in Eq.~\eqref{eq:spinup} overcomes the $\Gamma_-$ term, and the system simply relaxes to $\boldsymbol{L} = L_z \boldsymbol{z}$, $\boldsymbol{Q} = 0$. What we have just described is the pendulum evolution associated with the left-handed collisional instability.

In short, if $\sigma (0) > 0$, the system is expected to experience first the spin-reversing right-handed instability [Eq.~\eqref{eq:riccatisoln}] and then the spin-magnifying left-handed instability [Eq.~\eqref{eq:spinup}]. Under ideal conditions, the first of these conserves the length $Q$ and angular momentum $L$ of the pendulum. The second conserves neither. The left half of Fig.~\ref{numerical} shows numerical results consistent with this analysis. We have confirmed numerically that a system with $\sigma (0) < 0$ shows only the left-handed instability.

(For clarity, we reiterate that we are assuming $S_z > 0$ and $\Gamma_- > 0$ throughout this work. Statements involving the sign of $\sigma$ should be interpreted with this in mind.)

The third way to satisfy $\left\langle \boldsymbol{D} \times \boldsymbol{S} \right\rangle \cong \left\langle \boldsymbol{D} \right\rangle \times \left\langle \boldsymbol{S} \right\rangle$ is to have at least one of $\boldsymbol{S}$ and $\boldsymbol{D}$ vanish. In supernovae and mergers, the relevant case is $\boldsymbol{D} \cong 0$. This initial condition corresponds to the resonance-like (or zero-spin) instability, which has an enhanced growth rate proportional to $\sqrt{\mu S_z \Gamma_-}$ [Eq.~\eqref{eq:Omega0}]. A discussion of this final regime will complete our analysis of the collisional flavor pendulum.

Once the evolution commences, $D_T$ immediately becomes nonzero due to the $- \Gamma_- \boldsymbol{S}_T$ term in the second of Eqs.~\eqref{eq:SDEOMs}. The $\mu \boldsymbol{D} \times \boldsymbol{S}$ term in the first of Eqs.~\eqref{eq:SDEOMs} then causes $\boldsymbol{S}$ to precess with frequency $\mu D_T$ around $\boldsymbol{\hat{D}}_T$. The precession makes $S_T$ grow larger, which feeds back on the development of $D_T$. Now translating this narrative into equations, we approximate the development of $\boldsymbol{S}$ using
\begin{equation}
\dot{\boldsymbol{S}} = \omega (t) \boldsymbol{\hat{D}}_T \times \boldsymbol{S}, ~~ \omega (t) \equiv \mu D_T (t), \label{eq:Dz0Sdot}
\end{equation}
and the development of $\boldsymbol{D}_T$ using
\begin{equation}
\dot{\boldsymbol{D}}_T = - \Gamma_- \boldsymbol{S}_T, \label{eq:Dz0Ddot}
\end{equation}
neglecting the $- \Gamma_+ \boldsymbol{D}_T$ term because at the moment we are focused only on the early phase in which $D_T$ is small. From Eqs.~\eqref{eq:Dz0Sdot} and \eqref{eq:Dz0Ddot} we obtain
\begin{equation}
\omega (t) = \mu S \Gamma_- \int^t_0 dt' \left( - \boldsymbol{\hat{D}}_T (t') \right) \cdot \boldsymbol{\hat{S}}_T (t').
\end{equation}
The more closely aligned $\boldsymbol{\hat{S}}_T$ is with $- \boldsymbol{\hat{D}}_T$, the more rapidly $\boldsymbol{S}$ tilts and $D_T$ grows. Thus the resonance-like instability propels the system toward $\boldsymbol{\hat{D}} \cong - \boldsymbol{\hat{S}}$. This configuration is an attractor to which all three regimes lead. The expected pendulum motion in this case is a spiral from $+\boldsymbol{z}$ down along a direction in the $xy$-plane. The right half of Fig.~\ref{numerical} bears this out.

The collisional instability at $D_z \cong 0$ is unique both in its growth rate and in its relationship to coarse-graining. The other two regimes involve synchronization of $\boldsymbol{S}$ and $\boldsymbol{D}$ at least at the coarse-grained level. This one does not. The lack of synchronization is why the instability growth rate is enhanced by a factor of $\sim \sqrt{G_F n_\nu / \Gamma}$. It is also why the dynamics has finer, more complicated features (Fig.~\ref{numerical}).

\section{Discussion \label{sec:discussion}}

We have appealed to coarse-graining throughout this work. Now we step back and consider its significance from a more conceptual standpoint.

We derived an adiabatic solution [Eq.~\eqref{eq:riccatisoln}] of the pendulum equation of motion [Eq.~\eqref{eq:pendulum}] that conserves the energy and length of the pendulum. These qualities imply coherent evolution in flavor space. Coarse-graining is what makes coherent evolution a possibility. Collisional instability is always accompanied by the exchange of energy between the pendulum and its environment, but energy exchange at the fine-grained level still allows for conservative coarse-grained motion if the exchange takes the form of small-scale fluctuations. As demonstrated by the $\textrm{sgn}(\sigma) = +1 \rightarrow~\textrm{sgn}(\sigma) = -1$ pendulum inversion, coarse-grained motion can be large despite coarse-grained dissipation being negligible. The secular effect of the fluctuations is to adiabatically reverse the spin of the pendulum. The consequent motion is reversible---and yet, due to $\Gamma_-$, there is still an environment-associated asymmetry in the time evolution (Fig.~\ref{riccati}). Conservation of pendulum spin, unlike conservation of energy, is not restored even approximately by averaging over time.

The oscillation term $\mu \boldsymbol{D} \times \boldsymbol{S}$ has the potential to cause rapid dephasing of $\boldsymbol{S}$ and $\boldsymbol{D}$ in flavor space. If the dephasing is not sufficiently rapid, energy input from the environment builds up rather than fluctuating near zero. The extreme manifestation is the $\sigma < 0$ instability, in which $\boldsymbol{S}$ and $\boldsymbol{D}$ are locked in a fixed relative phase from the beginning. When the flavor evolution is coherent with respect to this phase, it is incoherent with respect to the flavor polarization.

The coherent limit of collisional instability---now going back to using \textit{coherent} to refer to flavor polarization---involves
\begin{equation}
\frac{\textrm{depolarization rate}}{\textrm{instability rate}} \sim \frac{\Gamma_+}{\Gamma_-} \frac{D_z}{S_z} \longrightarrow 0 \label{eq:coherent1}
\end{equation}
as well as
\begin{equation}
\frac{\textrm{fluctuation rate}}{\textrm{dephasing rate}} \sim \frac{\vartheta \Gamma_-}{\mu D_z} \ll 1, \label{eq:coherent2}
\end{equation}
where $\vartheta$ is the in-medium mixing angle. Equations~\eqref{eq:coherent1} and \eqref{eq:coherent2} ensure that the evolution occurs rapidly enough to avoid significant depolarization, but not so rapidly as to sacrifice the dephasing that is required for reversible motion. To satisfy Eq.~\eqref{eq:coherent1}, it is necessary to take $D_z / S_z \rightarrow 0$ because $\Gamma_+ > \Gamma_-$ always holds. However, this limit must be taken without entering the regime of the resonance-like collisional instability, in which the large $\Gamma_-$-driven fluctuations of $\boldsymbol{D}_T$ cause desynchronization.

At this point it might be helpful for us to attend to an apparent contradiction in the derivation of Eq.~\eqref{eq:pendulum}. In Eq.~\eqref{eq:STDTzero} we assumed that $\left\langle \boldsymbol{S}_T \right\rangle \cong 0$. The issue is that, by the definition of $\boldsymbol{Q}$, we seemingly must have $\boldsymbol{Q}_T \cong 0$, which implies that the pendulum is stuck at its initial position. The resolution is that we are using Eq.~\eqref{eq:STDTzero} to motivate approximate coarse-grained equations that capture the important features of the true coarse-grained dynamics. Once we are in possession of those equations, we no longer need to impose $\left\langle \boldsymbol{S}_T \right\rangle \cong 0$. Or, to put it differently, we no longer need to make reference to fine-grained variables at all.

We can restore consistency at an even deeper level by considering an ensemble of solutions $\boldsymbol{q}(t)$ with initial conditions $\boldsymbol{q}(0)$ that are uniformly distributed in the azimuthal coordinate. The ensemble-averaged $\boldsymbol{Q}_T$ is indeed zero, and the ensemble-averaged $\boldsymbol{q}(t)$ has no fine-grained dependence on initial conditions.

More profoundly, ensemble averaging connects our analysis to neutrino quantum thermodynamics \cite{johns2023c}. By the ergodic hypothesis, we can replace time averages by ensemble averages. The pendulum inversion of Eq.~\eqref{eq:riccatisoln} thus becomes an ensemble moving adiabatically through a sequence of equilibria.

Our crucial assumption that there are no fine-grained correlations [Eq.~\eqref{eq:ergodic}] is the statement that fluctuations go to zero in the thermodynamic limit. In Sec.~\ref{sec:pendulum} we made a distinction between the right-handed and left-handed instabilities based on the reason for Eq.~\eqref{eq:ergodic} being valid: respectively, alignment of $\boldsymbol{\hat{S}}$ and $\boldsymbol{\hat{D}}$ on the coarse-grained or (more restrictively) the fine-grained level. These conditions correspond to the system being in equilibrium or (again, more restrictively) a steady state.

We also noted that the case $D_z \cong 0$ is exceptional in that alignment is immediately lost due to the development of nonzero $\boldsymbol{D}_T$. Thermodynamically, it is exceptional because it begins at an unstable equilibrium and is immediately driven far from equilibrium by fluctuations. By contrast, the other regimes maintain alignment throughout their evolution and therefore experience no heating due to oscillations. They are kinematically adiabatic, in the language of Ref.~\cite{johns2023c}. In the coherent limit specified by Eqs.~\eqref{eq:coherent1} and \eqref{eq:coherent2}, the right-handed instability is fully adiabatic---no heating from kinematic or environmental decoherence---which is why it is reversible.

To close this study, we observe that there are some obvious and important extensions to consider. The model defined in Sec.~\ref{sec:model} is based on numerous simplifying assumptions, each of which could be relaxed. Introducing vacuum oscillations might be especially interesting, as the collisional flavor pendulum would then find itself in a gravitational field.

\begin{acknowledgments}
We warmly acknowledge conversations with Huaiyu Duan, Chinami Kato, Hiroki Nagakura, Meng-Ru Wu, and Zewei Xiong, and offer special gratitude to Wick Haxton and the rest of the N3AS community. During the completion of this work, LJ was supported as an Einstein Fellow through NASA grant number HST-HF2-51461.001-A and subsequently as a Feynman Fellow through LANL LDRD project number 20230788PRD1. SR was supported by the N3AS Physics Frontier Center through NSF award number 2020275.
\end{acknowledgments}

\bibliography{all_papers}

\end{document}